\documentclass[%
 reprint,
 amsmath,amssymb,
 aps,
]{revtex4-1}

\usepackage{graphicx}
\usepackage{dcolumn}
\usepackage{bm}

\begin{document}

\preprint{APS/123-QED}

\title{Stochastic Pore Collapse Models in Granular Materials}

\author{Joseph Bakarji}
\author{Daniel Tartakovsky}%
 \email{tartkovsky@stanford.edu}
\affiliation{%
Department of Energy Resources Engineering \\ 367 Panama Mall, Stanford University, Stanford, CA 94305
}%

\date{\today}

\begin{abstract}
Stochastic models for pore collapse in granular materials are developed. First, a general fluctuating stress-strain relation for a plastic flow rule is derived. The fluctuations account for non-associativity in plastic deformations typically observed in heterogeneous materials. Second, an axisymmetric spherical shell compaction model is extended to account for fluctuations in the material microstructure due to granular interactions at the pore scale. This changes the stress-strain constitutive equation determining the dynamics of pore collapse. Results show that stochastic differential equations can account for multiscale interactions in a statistical sense.

\end{abstract}

\maketitle


\section{Introduction}

Permanent deformations observed in heterogeneous media remain a major challenge in discrete-to-continuum multiscale modeling. The starting point in continuum modeling of a material's behavior is the identification of the appropriate constitutive relationship relating stress and strain, e.g., $\sigma_{ij} = \mu \dot \varepsilon_{ij}$ (linear viscous) or $\sigma_{ij} = \lambda \varepsilon_{ij}$ (linear elastic). This constitutive equation is then used in the Cauchy equation of motion $\rho D \mathbf u/Dt = \nabla \cdot \bm \sigma$ to describe the dynamical evolution of the velocity field constrained by the conservation of momentum. Accordingly, given well-posed initial and boundary conditions, the flow velocity $\mathbf u(\mathbf x,t)$ is uniquely determined.

The granular compaction problem is typically described by a set of hyperbolic conservation laws of mass, momentum and energy. There is a wide spectrum of assumptions that can be made. Due to the lack of accurate experimental data, this results in a large variety of models that depend on the physical regime, computational efficiency, and modeling tastes. Models can range anywhere from single phase Euler equations reaction equations \cite{Sen2018}, to two-phase equations with non-conservative compaction laws \cite{baer1986}, with different choices of thermodynamic closures, and stress-strain relations.

All these continuum models assume that subscale fluctuations do not affect the macroscopic dynamics, but this has been observed not to be true in energetic materials, especially when considering the fact that hotspot formation is due to extreme local thermal dissipation.
In granular media, even a ``simple'' Couette flow leads to jumps in the velocity profile, such as shear banding, showing onset of a chaotic behavior.
In such situations, increasing the complexity of a continuum model will not be enough to capture the behavior of the system itself.
Earlier work has shown that simply adding fluctuations to the initial microstructure accounts for the sensitivity of reaction-rate on the pore size \cite{bakarji2019microstructural}.
Motivated by the fluctuating Lifschitz-Landau fluctuating Navier-Stokes equation, we posit that one can account for unpredictable microscale phenomena in granular materials by introducing a fluctuation term into the continuum equations.

\begin{figure}[htbp]
\begin{center}
\includegraphics[width=4cm]{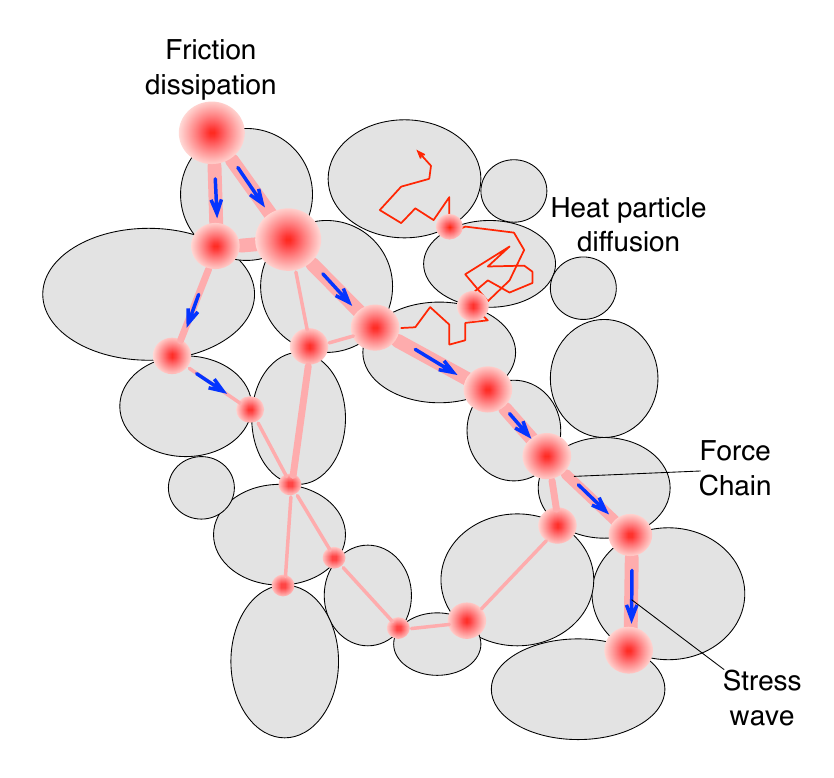}
\includegraphics[width=3cm]{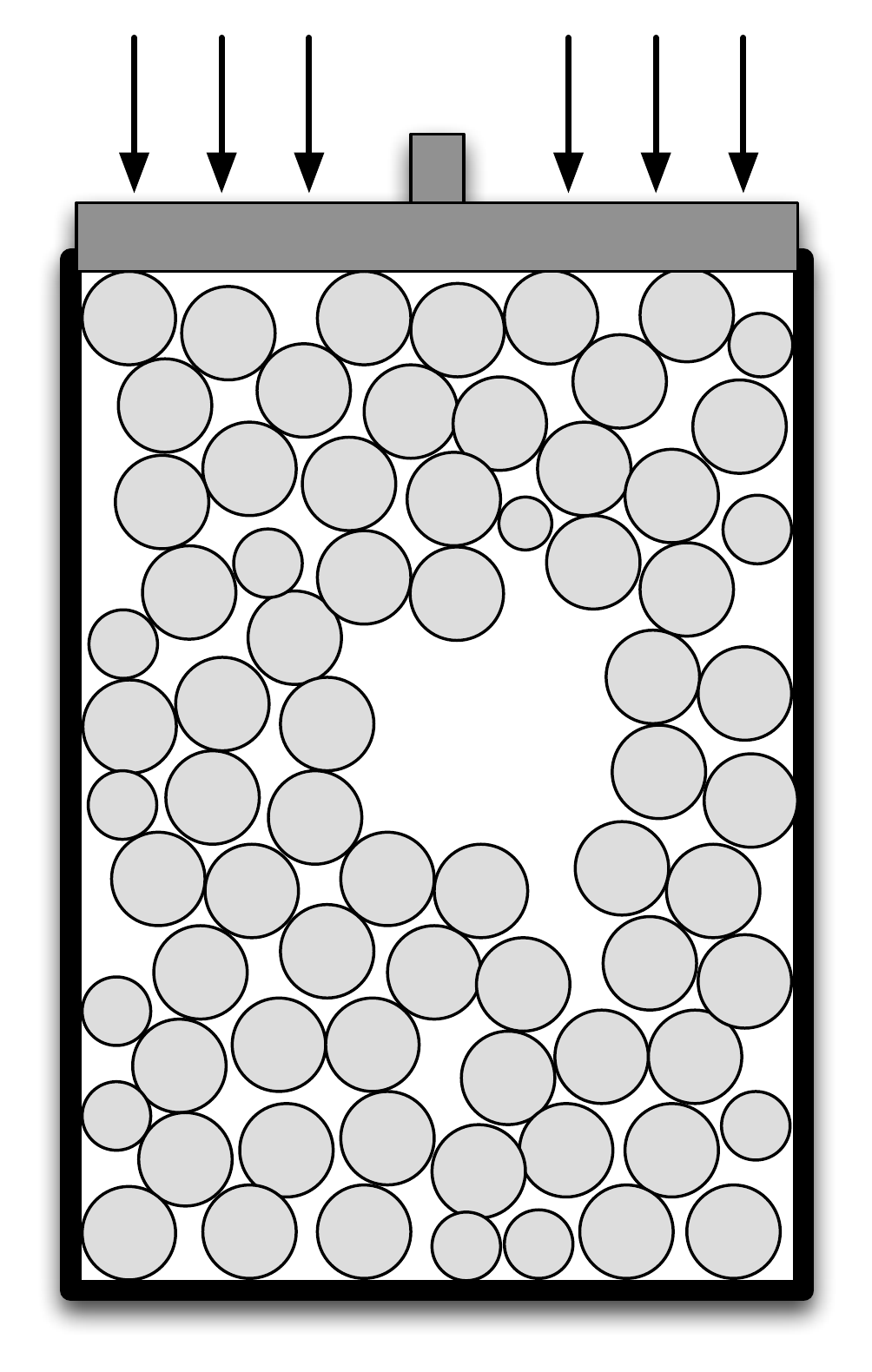}
\caption{{\small Interaction between granular dissipation due to friction, stress waves and heat diffusion.}}
\label{grangraph}
\end{center}
\end{figure}

A simple single phase continuum model accounts for conservation of mass, momentum and energy while being flexible to the constitutive relation

\begin{align*}
\text{Mass:  }& \frac{\partial \rho}{\partial t} + \frac{\partial \rho u_i}{\partial x_i} = 0\\
\text{Momentum:  }& \frac{\partial \rho u_i}{\partial t} + \frac{\partial (\rho u_i u_j + \tau_{ij})}{\partial x_j} = 0 \\ 
\text{Energy:  }& \frac{\partial \rho E}{\partial t} + \frac{\partial (\rho E u_j + \tau_{ij} u_i)}{\partial x_j} = \dot \epsilon
\end{align*}
With density $\rho$, velocity $u_i$, stress tensor $\tau_{ij}$, total energy $E$, and thermal dissipation $\dot \epsilon$.
An equation that models the evolution of the compaction is usually added.
For example \cite{Sen2018} uses the Prantdl-Ruess elastoplastic flow rule for the deviatoric stress tensor $s_{ij} = \tau_{ij} - \tau_m \delta_{ij}$ given by
\begin{equation}\label{prflowrule}
	\frac{\partial \rho s_{ij}}{\partial t} + \frac{\partial \rho s_{ij} u_k}{\partial x_k}  + \frac{2}{3} \rho G \dot \varepsilon_{kk} \delta_{ij} - 2 \rho G \dot \varepsilon_{ij} = 0
\end{equation}
Where $\dot \varepsilon_{ij}$ is the strain rate tensor and $G$ is the shear modulus of the material.

A simpler plastic deformation flow rule can be used, such as the $J_2$ flow potential (\cite{lubliner2008plasticity} pg. 114)
\begin{equation}
\dot \varepsilon^p_{ij} = \frac{1}{2\eta}  \left\langle 1 - \frac{k}{\sqrt{J_2}} \right\rangle s_{ij}
\end{equation}
where $\eta$ is a temperature-dependent viscosity, the Macauley bracket defined as $\left\langle x \right\rangle = x H(x)$, where $H$ is the heaviside step function, and $J_2 = 1/2 s_{ij} s_{ij} = 1/6[(\tau_1 - \tau_2)^2 + (\tau_2 - \tau_3)^2 + (\tau_3 - \tau_1)^2]$ \cite{lubliner2008plasticity}.

\section{Fluctuating Plasticity}

A heterogeneous microstructure introduces fluctuations in the stresses which introduce random slip planes in plastic deformations.
The corresponding uncertainty in the model can be accounted for by adding a stochastic term to the stress-strain constitutive relation.
Similarly to hydrodynamic fluctuations \cite{landau1981statistical}, a white noise $\xi(x, t)$ with zero mean and variance $\sigma_\xi^2$ is added to the flux such that

\begin{align*}
	\mathbb{E}[\xi(x, t)] &= 0 \\
	\mathbb{E}[\xi(x, t), \xi(y, \tau)] &= \sigma_\xi^2 \delta(x - y) \delta(t - \tau)
\end{align*}
Where the strength of the noise $\sigma_\xi^2$ is determined either (i) analytically via the fluctuation-dissipation theorem or (ii) numerically by upscaling discrete element method (DEM) simulations. In this study we focus on using the first method.

\emph{Why the fluctuations are not spatially correlated? Because the granular fluctuation time scale $\tau_f$ due to temperature and friction is assumed to be much faster than the time scale of the compaction $\tau_c$, that is $\tau_c \gg \tau_f$.}

Plasticity theory has been the conventional method for describing large deformations in amorphous particulate media, in particular granular media. 
However, one of the main assumptions usually used in describing plastic deformations is the associated flow rule, even though large deviations from this assumption were clearly observed in amorphous materials \cite{forterre2008flows}. 
This motivates more general, herein statistical, considerations for strain-increments that reflect the heterogeneity in the microstructure of granular media and capture the multiscale effects of local fluctuations.

For a given temperature $T$, an elasto-plastic material can be described by a yield surface, $\mathcal S(\boldsymbol \tau, T) = 0$, a corresponding plastic potential $g(\bm \tau, T)$, and a flow rule (~\cite{lubliner2008plasticity} pg. 114)
\begin{equation}\label{flowrule}
  \dot{\varepsilon}^p_{ij} = \lambda \frac{\partial g}{\partial \tau_{ij}}
\end{equation}
where $\dot{\varepsilon}^p_{ij}$ is the plastic part of the strain rate tensor $\dot{\bm\varepsilon} = \dot{\bm\varepsilon}^p + \dot{\bm\varepsilon}^e$. A deformation is purely elastic when $\mathcal S(\bm \tau,\cdot) < 0$ and is irreversibly plastic for $\mathcal S \ge 0$. A common assumption, especially for visco-plastic materials, is that the plastic potential is the same as the yield surface, such that $\partial_{\bm \tau} \mathcal S \propto \partial_{\bm \sigma} g$.

\begin{figure}[h]
\begin{center}
\includegraphics[width=7.8cm]{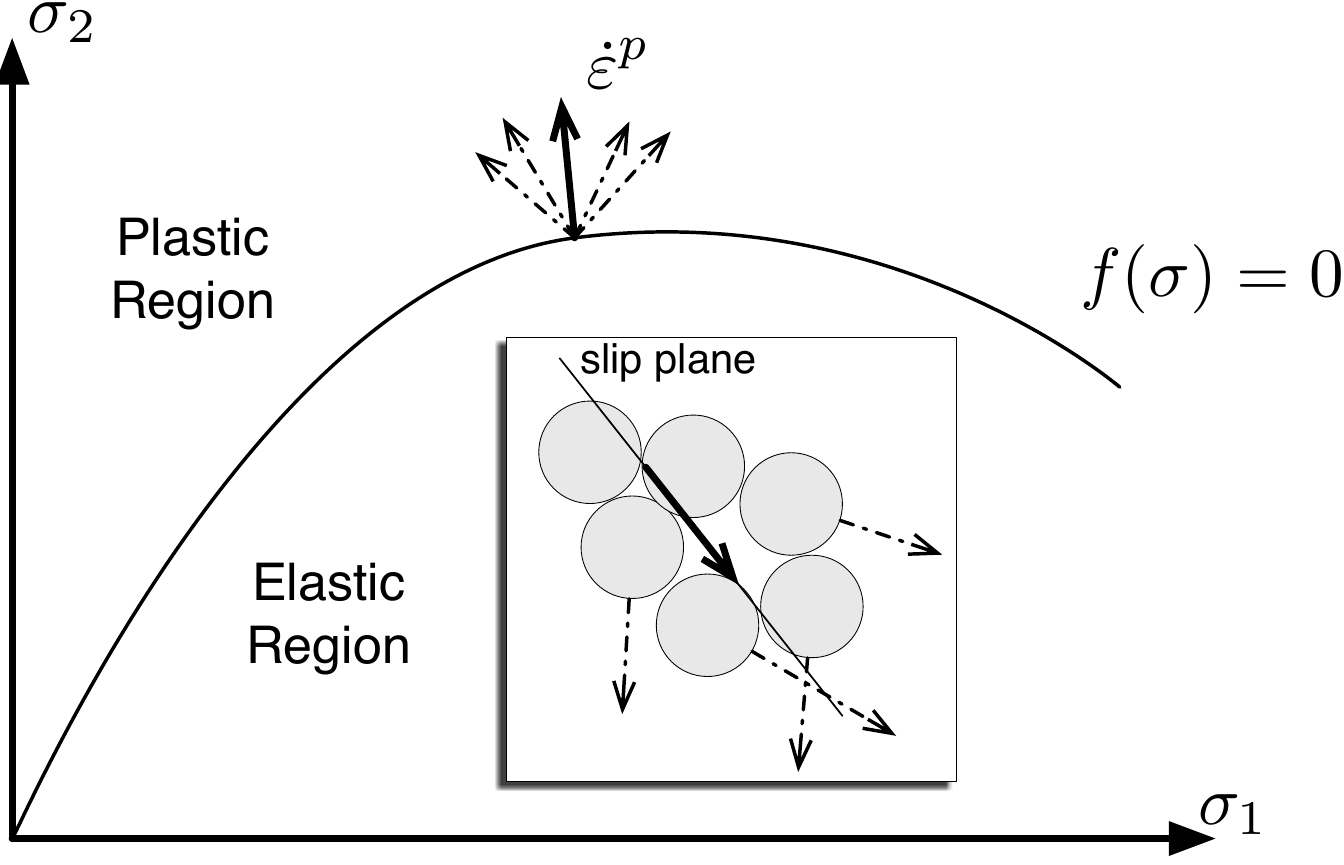}
\vspace{-3mm}
\caption{{\small The stochastically-associative flow rule.}}
\label{stochflowrule}
\end{center}
\end{figure}
This so-called \emph{normality} condition implies that the strain rate $\dot{\bm\varepsilon}^p$ is perpendicular to the yield surface pointing outward as shown in Fig. \ref{stochflowrule}. This assumption holds in metals but has had limited success in amorphous materials, such as granular media. 
In essence, yield occurs whenever a general measure of the local shear stress $\tau_s$ exceeds the normal stresses $\tau_n$; by analogy to the Mohr-Coulomb criterion.
Therefore, at the grain scale this criterion depends on the granular structure with $\tau_s$ and $\tau_n$ depending on the normal ($\mathbf F^i_n$) and tangential ($\mathbf F^i_t$) forces between the grains $i$.
This grain-scale inhomogeneity induces macroscopic effects, thus requiring a statistical description of yield surfaces at the grain level. We propose to do so by introducing the fluctuating tensor $\bm \xi$ into the flow rule,
\begin{equation}
	\dot \varepsilon_{ij}^p = \lambda \frac{\partial \mathcal{S}}{\partial \tau_{ij}} + \sigma_\xi \xi_{ij}(t)
\end{equation}

where the noise strength (variance) $\sigma_\xi$ depends on the microscopic variables such as grain size and surface roughness.
Since $\xi_{ij} = \text d \mathcal W_{ij} / \text dt$ where $\mathcal W$ is the Wiener process, the flow rule~\eqref{flowrule} takes the form of a Langevin equation 
\begin{equation*}\label{gransde}
  \text d\varepsilon^p_{ij} = \lambda \frac{\partial \mathcal{S}}{\partial \tau_{ij}} \text dt + \sigma_\xi \text d\mathcal W_{ij}.
\end{equation*}
In words, the strain associated with permanent deformations follows a Brownian path with a drift term guided by the slip plane (Fig.~\ref{stochflowrule}).
This is consistent with the fact that the displacement of individual grains is very sensitive to initial conditions and can only be known statistically at each time step.

To further motivate our approach, we look at the physical basis of stochastic plasticity.
According to Drucker's postulate (\cite{lubliner2008plasticity} pg. 125), a granular medium in compression is assumed to be stable as it undergoes pure hardening. 
Stochasticity in the yield surface adds local concavities accounting for grain-level instabilities. We propose a probabilistic extension of the Drucker postulate of stability, $\dot{\bm\tau} \dot{\bm\varepsilon}^p \ge \bm 0$, in the form 
\begin{equation}
\dot{\bm\tau} \dot{\bm\varepsilon}^p  \ge \gamma \bm\xi
\end{equation}
where $\gamma$ is the strength (variance) of the fluctuations $\bm\xi$.
This allows for the possibility that the work done in a loading-unloading cycle, $\text d \varepsilon \text d\tau$, is only on average positive but sometimes negative.
This argument can be constructed by analogy to the Fluctuation Theorem, which extends the second law of thermodynamics to include a finite probability that entropy increases only on average~\cite{Ostoja-Starzewski2014}.
This leads to a more general approach to plasticity.
Namely, the yield surface needs not be completely smooth and the strain-increments are not unique; they are chosen from a set with a probability distributions.

\subsection{Stochastic Thermoelasticity}
The concept of plastic potential was introduced by Mises in 1928 by analogy to the elastic potential which has a similar form to Eqn.~\eqref{flowrule}.
Likewise, we propose a stochastic extension to thermoelasticity and apply it to plasticity.
The result is a stochastic differential equation that is similar to Eqn.~\ref{stochflowrule} but provides more information about the microscale fluctuations.

Given a deforming elastic body, energy conservation can be expressed as 
\begin{equation}\label{energy}
\rho \dot E = \rho r - \nabla \cdot \mathbf h + \dot \varepsilon_{ij}^p \dot \sigma_{ij}
\end{equation}
where $E$ is the internal energy, $r$ the heat generation, $\mathbf h$ the heat flux, and $\dot \varepsilon_{ij}^p \dot \sigma_{ij}$ is the deformation power. The entropy $\eta$ is assumed to be a function of $E$ and the strain $\bm\varepsilon$, i.e., $\eta = \eta(E, \bm \varepsilon)$.
The basic stress-strain relation of thermoelasticity is derived by first assuming the body to be isentropic, $\dot \eta = 0$, and adiabatic, $\rho r - \nabla \cdot \mathbf h = 0$, which leads to 
\begin{equation}\label{thermoelasticeq}
\sigma_{ij} = - T \rho \frac{ \partial \eta}{ \partial \varepsilon_{ij}}, \qquad T = \frac{\partial E }{ \partial \eta }.
\end{equation}

Now let's consider a system which is not ``perfectly'' adiabatic.
A thermoelastic body can have a fluctuating thermal energy $Q \equiv \rho \dot E - \sigma_{ij} \dot \varepsilon_{ij} = \sigma_\xi \xi(t)$ where $\xi$ is a white noise of strength $\sigma_\xi$.
Eqn.~\eqref{energy} and the chain rule applied to the isentropic condition yields
\begin{equation} \label{crentropy}
\frac{\partial \eta}{\partial E} \frac{\partial u}{\partial t} + \frac{\partial \eta}{\partial \varepsilon_{ij}}\frac{\partial \varepsilon_{ij}}{ \partial t} = 0
\end{equation}
Using equations \ref{energy} and \ref{crentropy}, we have
\begin{equation}\label{eq:wtf}
\left(  \sigma_{ij} + T \rho \frac{\partial \eta}{\partial \varepsilon_{ij}}  \right) \dot \varepsilon_{ij} = \sigma_\xi \xi(t).
\end{equation}
In the deterministic case, $\xi(t) \equiv 0$ and, since the resulting equality holds for all $\dot \varepsilon_{ij}$, one recovers Eqn.~\eqref{thermoelasticeq}. In the stochastic case, this argument cannot be made and $\dot \varepsilon_{ij}$ has to be accounted for by inversion, such that
\begin{equation} \label{stochthermoelastic}
\bm \sigma = - T \rho \partial_{\bm\varepsilon} \eta + \sigma_\xi  \dot{\bm \varepsilon}^{-1} \xi(t).
\end{equation}
Thus, the addition of microscale noise into the continuum description gives rise to the dependence of stress on the strain rate, which is absent in the perfectly adiabatic case.
Equation~\eqref{thermoelasticeq} is seldom used in practice because fluctuations in the entropy rarely affect the macroscopic behavior, unless a phenomenon like hotspots is involved.
The same form can be obtained using Helmholtz free energy, $\Psi = E - T\eta$, with the same approach.

We will use Eqn.~\eqref{eq:wtf} to determine $\sigma_\xi = \sigma_\xi(\dot{\bm\tau})$, the noise strength in our stochastic flow rule~\eqref{stochflowrule}. 

Eqn.~\eqref{stochflowrule} and~\eqref{eq:wtf} reveal a time-varying dependence of the stress on the stress rate.
This is inline with recently observed dependence of material response on input force~\cite{Hasan2015}.
A rigorous characterization of the noise strength $\sigma_\xi(\dot{\bm\tau})$ will shed light on the stability of a granular structure as a function of microscopic state variables.

Finally, the quantitative knowledge of $\sigma_\xi(\dot{\bm\tau})$ will allow us to close a Fokker-Planck equation for the stochastic differential equation~\eqref{gransde},
\begin{equation}
	\frac{\partial p}{\partial t} = - \frac{\partial}{\partial \varepsilon_{ij}} \left( \frac{\partial g}{\partial \tau_{ij}} p   \right) + \frac{1}{2} \sum_{i,j} \frac{\partial^2}{\partial \varepsilon^2_{ij}} ( \sigma_\xi(\tau_{ij})^2 p).
\end{equation}
It describes the temporal evolution of the strain probability of deformation, $p(\varepsilon_{ij}, t | \varepsilon^0_{ij}, t_0)$. This equation can be solved numerically using reverse Brownian motion locally in Langevin equivalent \cite{Bakarji2017}.

\section{A Fluctuating Pore Collapse Model}
An axisymmetric compaction of a viscoplastic spherical shell developed by Carroll and Holt \cite{Carroll1972} is one of the earliest models for pore collapse in energetic materials.
Nesterenko and Carroll later proposed an extension that takes into account the temperature \cite{nesterenko2013} Fig.~\ref{chmodel}.
Since then, thanks to the exponential increase in computational power, much more sophisticated computational models have been proposed offering a lot of insight into the complex mechanics of pore collapse. \cite{Sen2018} 
However, the increasing complexity does not necessarily reflect a physical system whose dynamics is still poorly understood.
In the present context of numerical modeling and simulation, Caroll-Holt's model is a good reduced order model with enough flexibility to account for the change in porosity for multiscale modeling.

The main limitation of the model is the assumption that the stress is equally distributed in all directions.
This is never the case in real pores, which can significantly affect hotspot formation predictions.
Furthermore, plastic deformation in the model is assumed to be always happening along a constant plane, whose direction in real granular materials is constantly fluctuating, as depicted in Fig.~\ref{stochflowrule}.

The equation of motion for a spherical geometry is \cite{nesterenko2013}
\begin{equation}
	\rho_s \ddot{r} = \frac{\partial \tau_{rr}}{\partial r} + \frac{2}{r}(\tau_{rr} - \tau_{\theta \theta})
\end{equation}
Where $\rho_s$ is the solid density.
To account for small scale fluctuations in stresses, we add a temperature-dependent fluctuating term $\xi(r, t)$ to the stress-strain relationship
\begin{align*}
	\tau_{rr} - \tau_{\theta \theta} &= Y + 2\eta \left( \frac{\partial v}{\partial r} - \frac{v}{r} \right) \\
	&= Y - 6\eta \frac{\dot r}{r} + \xi(r, t) = 2 \tau_s + \xi(r, t)
\end{align*}
Where $Y$ is the yield strength, and $\eta$ is the viscosity, and $\tau_s$ the shear stress.
With the problem illustrated in Fig.\ref{chmodel}, the boundary conditions $\tau_{rr}(r = a) = 0$, $\tau_{rr}(r = b) = P(t)$, $a(0) = a_0$ and $\dot a(0) = 0$.
The dependence of the inner pore radius $a$ on the applied pressure $P(t)$ can be written as 
\begin{equation}
P(t) = 2 \int_a^b \tau_s \frac{dr}{r} + P_\text{kin} (\ddot a, \dot a, a)
\end{equation}
The second term, $P_\text{kin}$ represent the inertia-dependent part of the pressure and determines the main difference between the quasistatic and total pressures
\begin{equation*}
P_\text{kin} (\ddot a, \dot a, a) = -\rho_s \left[ (a \ddot a + 2 \dot a^2 ) \left(1 - \frac{a}{b} \right) -\frac{1}{2} \dot a^2 \left( 1 - \frac{a^4}{b^4} \right) \right]
\end{equation*}

For temperature dependent yield strength and viscosity, $Y$ and $\eta$ are given by
\begin{align*}
    Y=& 
\begin{cases}
    Y_1 (1 - T/T_m),& T \le T_m\\
    0,              & T > T_m\\
\end{cases}\\
\eta &= \eta_m \exp \left[ B\left( \frac{1}{T} - \frac{1}{T_m} \right)\right]
\end{align*}
where $Y_1$ is the yield strength at zero temperature, $T_m$ is the melting temperature, $\eta_m$ is the viscosity of the melt, and $B$ is a constant (here set to $5765 K^{-1}$).

Integrating equation \ref{mom} from $a$ to $b$, we get
\begin{equation*}
P(t) = 2 \int_a^b (Y - 6 \eta \frac{\dot{r}}{r})\frac{dr}{r} + P_{kin} (\ddot a, \dot a, a) + 2 \int_a^b \frac{\xi(r,t)}{r} dr 
\end{equation*}
Or
\begin{equation}
	P(t) = P_\text{CH}(a, \dot a, \ddot a, t) + 2 \text{ln}\left(\frac{b}{a}\right) \sigma_\xi \mathcal{W}(0, b-a)
\end{equation}
Where $P_\text{CH}(a, \dot a, \ddot a, t)$ is the original integro-differential term in the deterministic Carroll-Holt model
\begin{align*}
    P(t) &= 2 \int_a^b \left[ Y_1 \left( 1-\frac{T}{T_m} \right) - 6\eta_m \text{exp}\left(\frac{B}{T} - \frac{B}{T_m} \right) \frac{\dot{r}}{r} \right] \frac{dr}{r}\\
& - \rho_s \left[(a \ddot{a} + 2 \dot{a}^2 )(1-\frac{a}{b} ) - \frac{1}{2} \dot{a}^2 \left( 1- \left(\frac{a}{b}\right)^4 \right) \right]
\end{align*}

\begin{figure}[htbp]
\begin{center}
	\includegraphics[width=5cm]{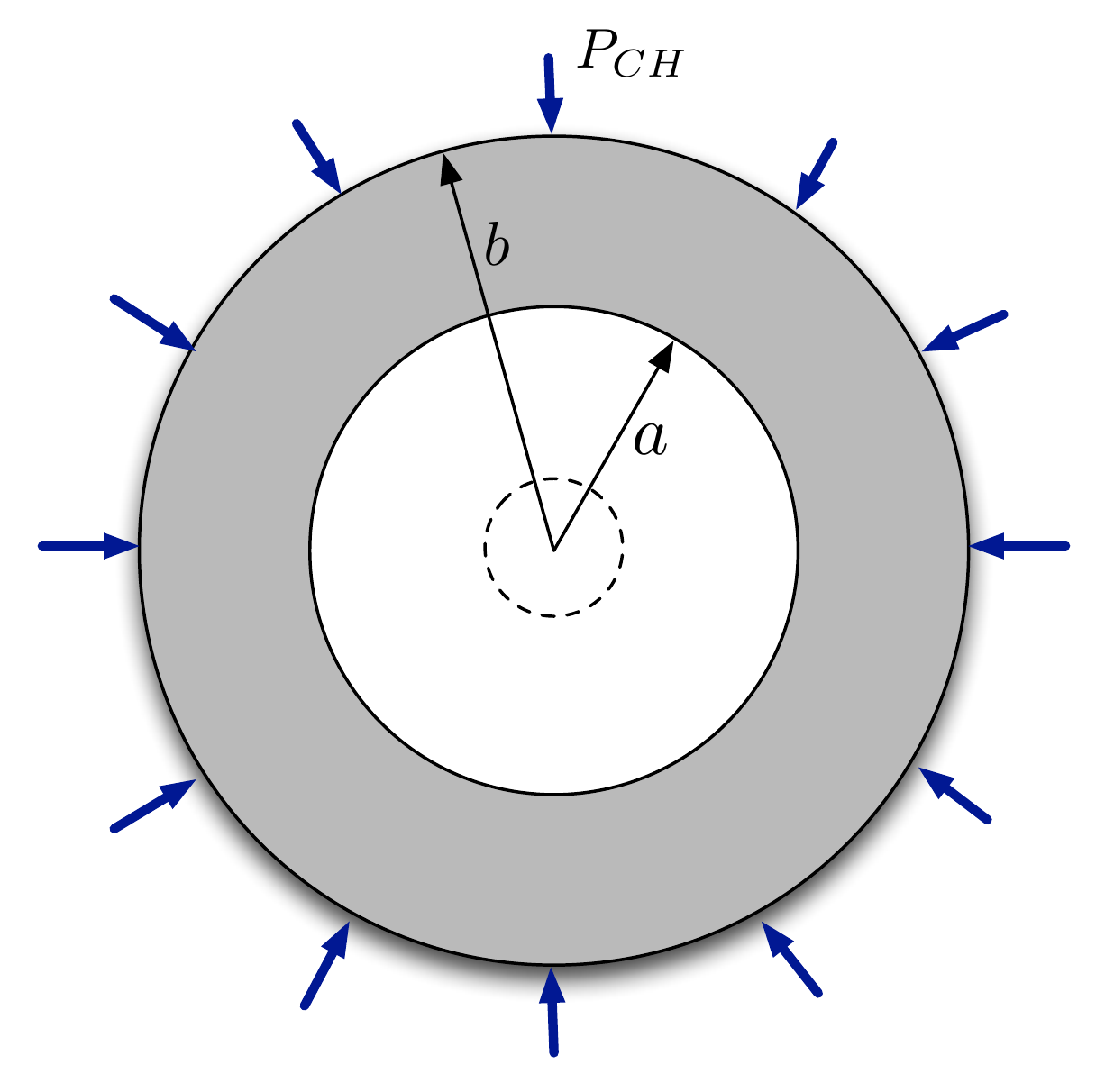}
\caption{Carroll-Holt pore collapse model}
\label{chmodel}
\end{center}
\end{figure}

The temperature is calculated using the conservation of energy $\partial E/\partial t = 2 \tau_s \dot \epsilon$ 
\begin{equation*}
\rho_s C_v \dot T = -2 Y_1 \left( 1 - \frac{T}{T_m} \right) \frac{\dot r}{r} + 12 \eta_m \exp \left[ \left( \frac{B}{T} - \frac{B}{T_m} \right) \frac{\dot{r}}{r} \right]
\end{equation*}
For $\eta = 0$, the temperature can be solved exactly as
\begin{equation}
T = T_m - (T_m - T_0) \left(\frac{r}{r_0} \right)^{\frac{2Y_1}{\rho_s C_v T_m}}
\end{equation}

\subsection{Fluctuation Strength}
To find the strength of the fluctuations, we use thermodynamic considerations as presented in Landau-Lifshitz fluctuating hydrodynamics \cite{landau1981statistical}.
In the case of a Navier-Stokes equation, the fluctuating terms $s_{ik}$ and $g_i$ are added to the stress $\sigma_{ik}$ and Fourier heat flux $\mathbf q = \kappa \nabla \partial T$ respectively.
Having an expression for the change in entropy $\dot S$ from fluid mechanics, the corresponding strength of the noise can be deduced.
These formulas are used

\begin{equation}
\dot S = \int \left[ \frac{1}{2T} \sigma_{ij} \left( \frac{\partial v_i}{\partial x_k} + \frac{\partial v_k }{\partial x_i} \right) - \frac{\mathbf q \cdot \nabla T}{T^2} \right] dV
\end{equation}
Defining $X_i = - \partial S/\partial x_i$ as the thermodynamic conjugate of a quantity $x_i$, we would like to find $\gamma_{ij}$ such that 
\begin{equation}\label{stochdec}
\dot x_i = - \sum_j \gamma_{ij} X_j + y_j
\end{equation}
where $y_j$ is a fluctuation term around the mean. By comparison, we find that 
\begin{table}[h!]
\begin{tabular}{ccc}
$\dot x_a$ = & $\sigma_{ik}$ & $q_i$ \\
$X_a$ = & $-1/(2T) \dot \epsilon_{ik} \Delta V$ & $1/T^2 \nabla T \Delta V$ \\
$\dot y_a$ = & $s_{ik}$ & $g_i$\\
$\gamma_{ij}$ = & $2T\eta A$ + $(\zeta - 2/3 \eta) B$ & $2\kappa T^2$ 
\end{tabular}
\end{table}

Where the last row was derived to be consistent with the relationship Eqn.~\ref{stochdec}. A is the shear dependent term, and (B) is the compressible term (See appendix~\ref{llfh} for more details)
Given the absence of heat flux $\bm q$ in the energy equation of the CH model, the fluctuations will only be revealed in the stress-strain relation.

The fluctuation strength $\sigma_\xi$ depends on the granular microstructure and temperature from thermodynamic considerations 
\begin{align*}
	\sigma_\xi^2(r) &= \int_a^b 2 T \eta(T) dr \\
	 &= \eta_m \int_a^b \left( T_m - (T_m - T_0) \left( \frac{r}{r_0}\right)^{\frac{2 Y_1}{ \rho_s C_v T_m}} \right) \times \\
	 &\exp{ \left(\frac{B}{T_m - (T_m - T_0) \left( \frac{r}{r_0}\right)^{\frac{2 Y_1}{ \rho_s C_v T_m}}} - \frac{B}{T_m} \right)}dr + \sigma_{\xi_m}^2
\end{align*}
Where $T_m$ is the melting temperature, $\eta_m$ the corresponding melting viscosity, $B$ a constant that depends on the material (in $K^{-1}$), $C_v$ the specific heat capacity, $\rho_s$ the solid density, and $\sigma_{\xi_m}$ the material-dependent thermal noise at melting temperature.

\begin{figure}[htbp]
\begin{center}
	\includegraphics[width=8cm]{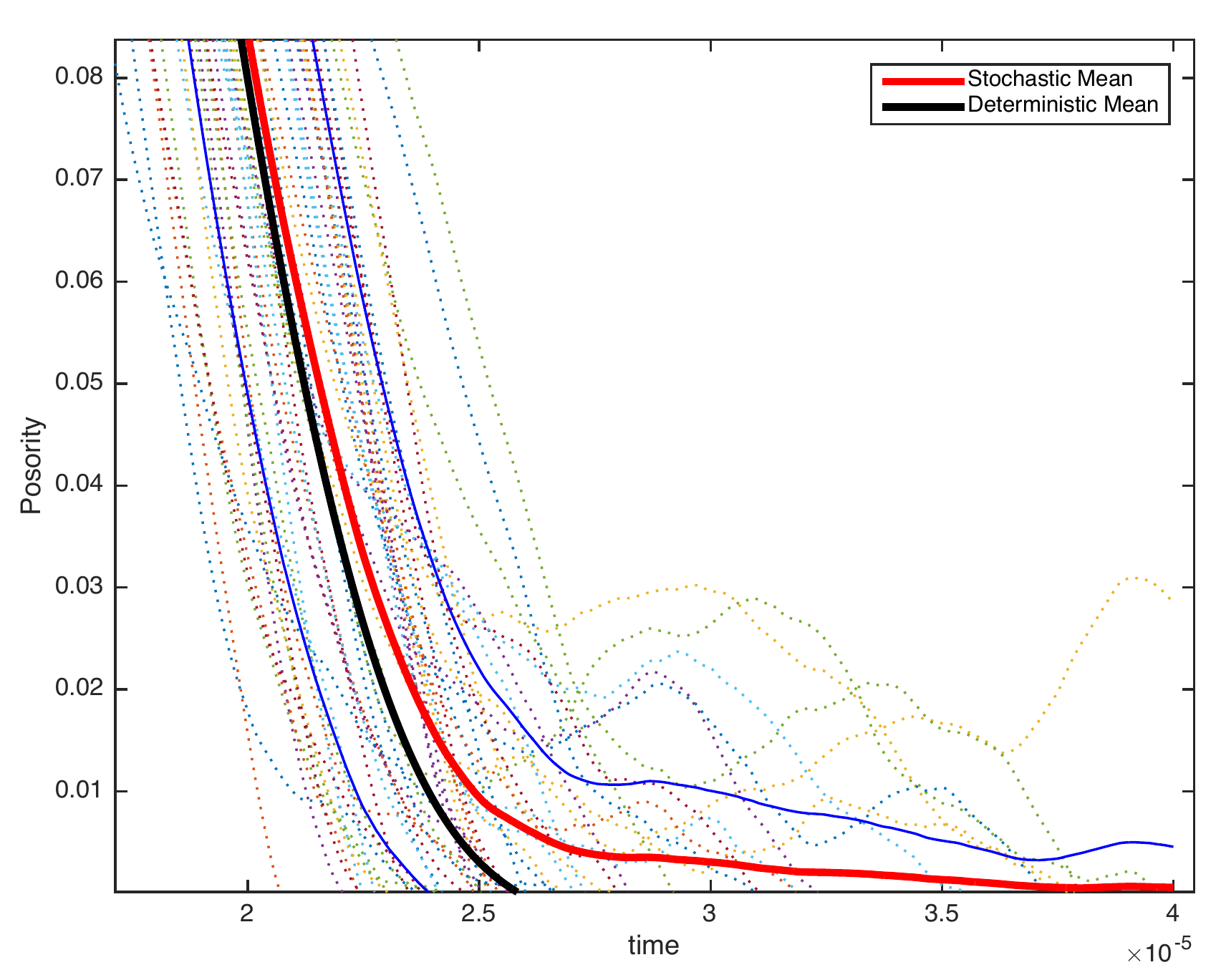} \\
	\includegraphics[width=8cm]{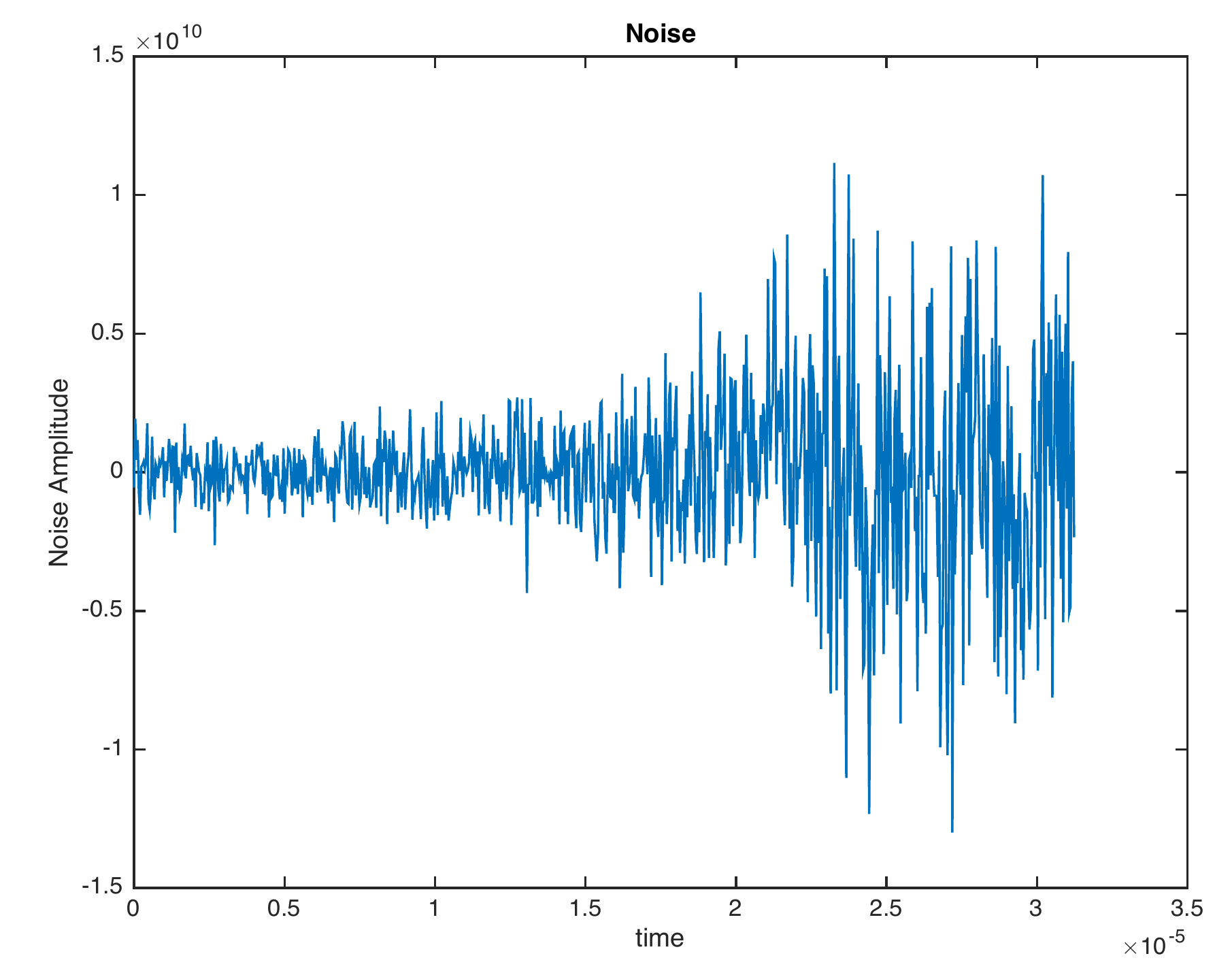}
\caption{The noise added in this figure is exaggerated to show the difference between mean and variance}
\label{chmodel}
\end{center}
\end{figure}

\section{Conclusion}
In this study we introduced two mathematical frameworks for modeling mesoscale pore collapse in compacted granular materials. The hard constraint of the associativity in plasticity can be relaxed by adding fluctuations to the flow rule. Similarly, fluctuations can be added to the stress-strain relation of the Carroll-Holt pore collapse model to relax its axisymmetric assumption. We've shown that this approach can account for the random microstructural fluctuations of the granular material at hand without explicitly tracking their deformations. We believe that this modeling procedure has potential in other fields of heterogeneous materials.

\appendix
\section{Landau-Lifshitz Fluctuating Navier-Stokes}\label{llfh}
According to \cite{landau1981statistical}, the equations of hydrodynamics can be written as
\begin{gather*}
\frac{\partial \rho}{\partial t} + \nabla \cdot (\rho \mathbf v) = 0 \\
\rho \frac{\partial v_i}{\partial t} + \rho v_k \frac{\partial v_i}{\partial x_k} = - \frac{\partial P}{\partial x_i} + \frac{\partial \sigma'_{ij}}{\partial x_k} \\
\rho T \left( \frac{\partial s}{\partial t} + \mathbf v \cdot \nabla s \right) = \frac{1}{2} \sigma'_{ik} \left( \frac{\partial v_i}{\partial x_k} + \frac{\partial v_k}{\partial x_i} \right) - \nabla \cdot \mathbf q
\end{gather*}
with no specific form of the stress tensor $\sigma'_{ij}$ and the heat flux vector $\mathbf q$.
For a compressible fluid, the constitutive equation and heat flux are expressed as
\begin{align*}
\sigma'_{ij} &= \eta \left( \frac{\partial v_i}{\partial x_k} + \frac{\partial v_k}{\partial x_i} - \frac{2}{3} \delta_{ik} \nabla \cdot \mathbf v \right) + \zeta \delta _{ik} \nabla \mathbf v + s_{ik} \\
\mathbf q &= - \kappa \nabla T + \mathbf g
\end{align*}
where $\eta$ and $\zeta$ are the viscosity coefficients, and $\kappa$ the thermal conductivity. The problem is to establish the properties of $s_{ik}$ and $\mathbf g$ as regards to their correlation functions.

Landau and Lifshitz \cite{landau1981statistical} show that 
\begin{gather*}
\langle s_{ik}(t_1, \mathbf r_1) g_l(t_2, \mathbf r_2) \rangle = 0\\
\langle g_i(t_1, \mathbf r_1) g_k(t_2, \mathbf r_2) \rangle = 2 \kappa T^2 \delta_{ik} \delta(\mathbf r_1 - \mathbf r_2) \delta(t_1 - t_2)\\
\langle s_{ik}(t_1, \mathbf r_1) s_{lm}( t_2, \mathbf r_2) \rangle = \\2T[ \eta(\delta_{il}\delta_{km} + \delta_{im} \delta_{kl} ) + (\zeta - 2/3 \eta) \delta_{ik} \delta_{lm}] \delta(\mathbf r_1 - \mathbf r_2) \delta(t_1 - t_2)
\end{gather*}

\newpage
\bibliography{energeticbiblio, energeticbiblio_extra}

\end{document}